\documentclass[aps,showpacs,pre,superscriptaddress]{revtex4}
\newcommand{\be}{\begin{equation}}
\newcommand{\ee}{\end{equation}}
\newcommand{\bea}{\begin{eqnarray}}
\newcommand{\eea}{\end{eqnarray}}
\newcommand{\sn}{{\rm sn}}

\newcommand{\dn}{{\rm dn}}
\newcommand{\cn}{{\rm cn}}
\newcommand{\sech}{{\rm sech}}

\vspace{.5in}
\usepackage{epsfig}
\begin{document}
\title{Exact Solutions of the Two-Dimensional Discrete Nonlinear Schr{\"o}dinger 
Equation with Saturable Nonlinearity}
\author{Avinash Khare}
\address{Institute of Physics, Bhubaneswar, Orissa 751005, India}
\author{Kim~{\O}. Rasmussen}
\address{Theoretical Division and Center for Nonlinear Studies,
Los Alamos National Laboratory,
Los Alamos, New Mexico, 87545, USA}
\author{Mogens R. Samuelsen}
\address{Department of Physics, The Technical University of Denmark,
DK-2800 Kgs. Lyngby, Denmark}
\author{Avadh Saxena}
\address{Theoretical Division and Center for Nonlinear Studies, Los 
Alamos National Laboratory, Los Alamos, New Mexico, 87545, USA}

\date{\today}

\begin{abstract}
We show that the two-dimensional, nonlinear Schr{\"o}dinger lattice with 
a saturable nonlinearity admits periodic and pulse-like exact solutions. We 
establish the general formalism for the stability considerations of these 
solutions and give examples of stability diagrams. Finally, we show that the 
effective Peierls-Nabarro barrier for the pulse-like soliton solution is zero.
\end{abstract}
\maketitle

\section{\bf Introduction.}

The discrete nonlinear Schr\"odinger equation (DNLSE) finds widespread use 
in physics due to its very general nonlinear character. It arises in the context of
the propagation of electromagnetic waves in optical waveguides
\cite{emwave}, and it also appears in the study of Bose-Einstein condensates in 
optical lattices \cite{bec}.  Recently we have studied the exact soliton
solutions and their stability for the one-dimensional DNLSE with a saturable
nonlinearity \cite{krss1}.   We were also able to obtain staggered and short
period solutions of this equation \cite{krss2} as well as to generalize our
results to arbitrarily higher-order nonlinearities \cite{krsss}.   

Two-dimensional periodic lattices with a saturable nonlinearity in the Schr\"odinger 
equation have been experimentally realized in photorefractive materials \cite{efrem}. 
Solitons have also been observed in these crystals \cite{fleischer,martin}.   Localized 
traveling wave solutions that exist only for finite velocities have been computed in 
this case \cite{melvin}.   The question of discrete soliton  mobility in these systems has 
been addressed as well \cite{magnus}.  Here we show that it is also possible  to 
obtain exact periodic and pulse-like soliton solutions for the two-dimensional DNLSE 
with a saturable nonlinearity.  We then study the stability of these solutions as a 
function of the parameters of the equation.  We also show that similar to the one 
dimensional case, the effective Peierls-Nabarro barrier (i.e. the discreteness barrier for soliton motion) for the pulse-like soliton 
solutions is zero.  In addition, we find several short period solutions. 

\section{\bf Two-dimensional discrete nonlinear Schr\"odinger equation with
saturable nonlinearity.}

The equation we consider is the following asymmetric, DNLSE with a
saturable nonlinearity in two dimensions
\be\label{field EQ}
 i\frac{d\phi_{n,m}}{dt}+ [\zeta(\phi_{n+1,m}+\phi_{n-1,m})
+\xi(\phi_{n,m+1}+\phi_{n,m-1})
] +\frac{\nu |\phi_{n,m}|^2 \phi_{n,m}}{1+|\phi_{n,m}|^2} =0\,.
 \ee
$|\zeta-\xi|$ is a measure of the spatial asymmetry and $\nu$ a measure of the
nonlinearity. This equation can be derived
from the Hamiltonian
\bea\label{Hamiltonian}
&&H=\sum_{n,m=1}^{N,M} \bigg [\zeta|\phi_{n+1,m}-\phi_{n,m}|^2
+\xi|\phi_{n,m+1}-\phi_{n,m}|^2
\nonumber \\
&&-[2(\zeta+\xi)+\nu]|\phi_{n,m}|^2
+\nu \ln(1+|\phi_{n,m}|^2) \bigg ]\,,
\eea
and the equation of motion being derived from
\be\label{Poisson Brackets}
i\dot{\phi}_{n,m} =\frac{\partial H}{\partial\phi_{n,m}^*},
\ee
considering $\phi_{n,m}$ and $i\phi_{n,m}^*$ as conjugate variables.
There are two conserved quantities for the field equation, Eq.
(\ref{field EQ}), the Hamiltonian $H$ and the power (norm) $P$ defined by
\be\label{Norm}
P=\sum_{n,m=1}^{N,M} |\phi_{n,m}|^2\,.
\ee 
Note that the system is invariant under simultaneous interchange of $\zeta$
and $\xi$, and $n$ and $ m$.  
%****Equation (\ref{field EQ}) can be considered as M 
%identical one-dimensional systems described in Ref. \cite{krss1} with a linear 
%nearest neighbor coupling $\xi$.
%****
%Kim had agreed that some clarification is required here since this statement
%while may be correct in the context of our solutions, is not correct 
%in general.****

\section{\bf Exact solutions to the two-dimensional equation.}

Exact stationary solutions can also be obtained in
the case of this two-dimensional discrete, asymmetric saturable 
nonlinear Schr\"odinger equation (\ref{field EQ}).
We are looking for stationary solutions using the ansatz
\be\label{time dependence}
\phi_{n,m}(t)=u_{n,m} e^{-i(\omega t+\delta)}\,,
\ee
and obtain from Eq. (\ref{field EQ}), the following difference equation
\be\label{time independent}
(1+ u_{n.m}^2)\omega u_{n,m}+(1+ u_{n,m}^2)
[\zeta(u_{n+1,m}+u_{n-1,m})+\xi(u_{n,m+1}+u_{n,m-1})]+\nu u_{n,m}^3=0\,.
\ee
Following Ref. \cite{krss1} we immediately find two different
types of solutions. One that is symmetric in $n$ and $m$ and one that only
depends on $n$ (and by symmetry one that only depends on $m$).\\

\noindent{\it Symmetric case:}

If one chooses 
\be\label{omega symmetric}
\omega=-\nu\,,
\ee
then
\be\label{dn symm}
u_{n,m}^s=\frac{\sn(\beta,k)}{\cn(\beta,k)}\dn(\beta (n+m+\delta_1),k)\,,
\ee
is a solution if $k$ is chosen to fulfill
\be\label{dn symm cond}
\frac{\nu}{\zeta +\xi}=2\frac{\dn(\beta,k)}{\cn^2(\beta,k)}\,,~~
\beta=\frac{2K(k)}{N_p}\,.
\ee
Similarly
\be\label{cn symm}
u_{n,m}^s=k\frac{\sn(\beta,k)}{\dn(\beta,k)}\cn(\beta (n+m+\delta_1),k)\,,
\ee
is a solution provided
\be\label{cn symm cond}
\frac{\nu}{\zeta +\xi}=2\frac{\cn(\beta,k)}{\dn^2(\beta,k)}\,,~~
\beta=\frac{4K(k)}{N_p}\,.
\ee
Here $k$ is the elliptic modulus (the
elliptic parameter $m=k^2$ \cite{stegun}) of the Jacobi elliptic functions
$\sn(x,k)$, $\cn(x,k)$, and $\dn(x,k)$ and $K(k)$ is the complete elliptic
integral of the first kind \cite{stegun,Ryzhik}. The integer $N_p$ denotes
the spatial period of the system. $N$ and $M$ in Eq. (\ref{Hamiltonian})
must be chosen as multiples of $N_p$. The two solutions have a common pulse-like 
limit for $k\rightarrow 1$ (and $N_p \rightarrow \infty$),
\be\label{pulse symm}
u_{n,m}^s=\sinh(\beta)\sech[\beta (n+m+\delta_1)]\,,
\ee
which is a solution if $\beta$ fulfills
\be\label{pulse symm cond}
\frac{\nu}{\zeta+\xi}=2 \cosh(\beta)\,.
\ee
By symmetry, a change of sign of $n$ (or $m$) in Eqs. (\ref{dn symm}),
(\ref{cn symm}), and (\ref{pulse symm}) will give solutions with exactly
the same properties. \\

\noindent{\it Asymmetric case:}

If one chooses for the frequency
\be\label{omega asymmetric}
\omega=-\nu-2\xi\,,
\ee
then
\be\label{dn asymm}
u_{n,m}^{as}=\frac{\sn(\beta,k)}{\cn(\beta,k)}\dn(\beta (n+\delta_1),k)\,,
\ee
will be a solution for $k$ satisfying
\be\label{dn asymm cond}
\frac{\nu}{\zeta}=2\frac{\dn(\beta,k)}{\cn^2(\beta,k)}\,,~~
\beta=\frac{2K(k)}{N_p}\,.
\ee
Similarly, we also have a $cn$ solution.
\be\label{cn asymm}
u_{n,m}^{as}=k\frac{\sn(\beta,k)}{\dn(\beta,k)}\cn(\beta (n+\delta_1),k)\,,
\ee
provided
\be\label{cn asymm cond}
\frac{\nu}{\zeta}=2\frac{\cn(\beta,k)}{\dn^2(\beta,k)}\,,~~
\beta=\frac{4K(k)}{N_p}\,.
\ee
Here $N=N_p$ and $M$ can be any integer.\\
Again both these solutions approach the pulse solution in the limit
$k \rightarrow 1$ and $N_p \rightarrow \infty$
\be\label{pulse asymm}
u_{n,m}^{as}=\sinh(\beta)\sech[\beta (n+\delta_1)]\,,
\ee
provided
\be\label{pulse asymm cond}
\frac{\nu}{\zeta}=2 \cosh(\beta)\,.
\ee
Another asymmetric solution appears if we interchange $n$ and $m$ and $\zeta$
and $\xi$ ($M=N_p$ and $N$ any integer).  We note that, in all cases, 
changing the 
sign of $\xi$ is equivalent to staggering the solution
in the $m$ direction ($(-1)^m$ as an amplitude factor) and changing the sign
of $\zeta$ is equivalent to staggering the solution in the $n$ direction
\cite{krss2}. 

The described solutions are in some sense direct generalizations
of our earlier results for the one-dimensional version of the 
saturable nonlinear
Schr{\"o}dinger equation, since they remain spatially uniform along a specific
direction in space. Such solutions are well-known for nonlinear partial 
differential equations and are often, in such a continuum setting, referred 
to as line solutions or line solitons. However, in continuum settings, 
such solutions are generally not stable, because the extra dimension now 
allows for an entire set of new instability modes to come into play. In 
our discrete case, however, we shall demonstrate that these solutions 
indeed can be stable in certain cases. The fact that these solutions have infinite
extension along one of their dimensions probably renders them less physically important. However,
their stability analysis does, as we will demonstrate, give detailed insight into the intricate stability mechanisms of this 
nonlinear system. Specifically, are the parameters $(\nu, \xi)$, which control the stability, directly related to materials properties such as the change in refractive index
of the crystal. 

\section{\bf Stability of the solutions.}

In order to study the linear stability of these exact solutions $u_{n,m}^j$
($j$ is ``s" (symmetric) or ``as" (asymmetric)) we introduce the following expansion
around the exact solution
\be\label{mrs}
\phi_{n,m}(t)=u_{n,m}^je^{-i(\omega t+\delta)}
 +\delta u_{n,m}(t)e^{-i(\omega t+\delta)}
\ee
applied in a frame rotating with frequency $\omega$ of the solution. 
Substituting (\ref{mrs}) into the field equation, Eq. (\ref{field EQ}), and
retaining only terms linear in the deviation, $\delta u_{n,m}$, we get
\begin{equation}
i\delta\dot{u}_{n,m}+\zeta\big(\delta u_{n+1,m}+\delta u_{n-1,m}\big)
+\xi\big(\delta u_{n,m+1}+\delta u_{n,m-1}\big)+
\left(\omega +\frac{\nu|u_{n,m}^j|^2(2+|u_{n,m}^j|^2)}{(1+|u_{n,m}^j|^2)^2}\right)
\delta u_{n,m}+
\frac{\nu|u_{n,m}^j|^2}{(1+|u_{n,m}^j|^2)^2}\delta u_{n,m}^*
=0.
\label{EQ:STAB_LIN}
\end{equation}
We continue by splitting the deviations $\delta u_{n,m} $ into real parts
$\delta u_{n,m}^{(r)}$
and imaginary parts $\delta u_{n,m}^{(i)}$ ($\delta u_{n,m}
=\delta u_{n,m}^{(r)}+i\delta u_{n,m}^{(i)}$) and introducing the two real
vectors
\begin{eqnarray}
\delta\mbox{\boldmath $U^r$}
=\{\delta u_{n,m}^{(r)}\}
=\{\delta U_{J}^{(r)}\},&~~\mbox{and}~~&
\delta\mbox{\boldmath $U^i$}
=\{\delta u_{n,m}^{(i)}\}
=\{\delta U_{J}^{(i)}\},
\end{eqnarray}
where the pair of indices $m,n$ are replaced by a single
index $J$ via: $J=n+(m-1)N_p$. By introducing the real matrices $\mbox{\boldmath $A$}=\{A_{J,J'}\}$ and
$\mbox{\boldmath $B$}=\{B_{J,J'}\}$ defined by 
\begin{eqnarray}
A_{J,J'}=
\zeta(\delta_{J,J'+1}+\delta_{J,J'-1})+
\xi(\delta_{J,J'+N_p}+\delta_{J,J'-N_p})
+ \left(\omega+\frac{\nu|u_{n,m}^j|^2(3+|u_{n,m}^j|^2)}
{(1+|u_{n,m}^j|^2)^2}\right)
\delta_{J,J'},\\
B_{J,J'}=
\zeta(\delta_{J,J'+1}+\delta_{J,J'-1})+
\xi(\delta_{J,J'+N_p}+\delta_{J,J'-N_p})
+\left (\omega+\frac{\nu|u_{n,m}^j|^2}{(1+|u_{n,m}^j|^2)}\right )
\delta_{J,J'}~~~~~~~~~~
\end{eqnarray}
%\begin{eqnarray}
%A_{nn'mm'}=
%\zeta(\delta_{n,n'+1}+\delta_{n,n'-1})\delta_{m,m'}+
%\xi\delta_{n,n'}(\delta_{m,m'+1}+\delta_{m,m'-1})
%+ \left(\omega+\frac{\nu|u_{n,m}^j|^2(3+|u_{n,m}^j|^2)}{(1+|u_{n,m}^j|^2)^2}\right)
%\delta_{n,n'}\delta_{m,m'},~~~~~\\
%B_{nn'mm'}=
%\zeta(\delta_{n,n'+1}+\delta_{n,n'-1})\delta_{m,m'}+
%\xi\delta_{n,n'}(\delta_{m,m'+1}+\delta_{m,m'-1})
%+\left (\omega+\frac{\nu|u_{n,m}^j|^2}{(1+|u_{n,m}^j|^2)}\right )
%\delta_{n,n'}\delta_{m,m'},~~~~~~~~~~~~~~~
%\end{eqnarray}
where  $J'\pm 1$ and $J'\pm N_p$ in the Kronecker $\delta$ means:
$J'\pm 1~mod~N_p$ and
$J'\pm N_p~mod~N_p$ to ensure periodic boundary conditions, 
Eq. (\ref{EQ:STAB_LIN}) %\cite{multiplication}
becomes
\begin{eqnarray}
-\delta\mbox{\boldmath $\dot{U^i}$}+
\mbox{\boldmath $A$}\delta\mbox{\boldmath $U^r$}=\mbox{\boldmath $0$},
&\mbox{and}&
\delta\mbox{\boldmath $\dot{U^r}$}+
\mbox{\boldmath $B$}\delta\mbox{\boldmath $U^i$}=\mbox{\boldmath $0$}.
\end{eqnarray}
Combining these first order differential equations we get:
\begin{eqnarray}
\label{eigenvalue}
\delta\mbox{\boldmath $\ddot{U^i}$}+\mbox{\boldmath $A$}\mbox{\boldmath $B$}
\delta\mbox{\boldmath $U^i$}=\mbox{\boldmath $0$},&\mbox{and}&
\delta\mbox{\boldmath $\ddot{U^r}$}+\mbox{\boldmath $B$}\mbox{\boldmath $A$}
\delta\mbox{\boldmath $U^r$}=\mbox{\boldmath $0$}.
\end{eqnarray}
The two matrices $\mbox{\boldmath $A$}$ and $\mbox{\boldmath $B$}$ are
symmetric and have real elements. However, since they do not commute
$\mbox{\boldmath $A$}\mbox{\boldmath $B$}$ and
$\mbox{\boldmath $B$}\mbox{\boldmath $A$}=
(\mbox{\boldmath $A$}\mbox{\boldmath $B$})^{tr}$
are not symmetric.
$\mbox{\boldmath $A$}\mbox{\boldmath $B$}$ and
$\mbox{\boldmath $B$}\mbox{\boldmath $A$}$ have the same eigenvalues,
but different eigenvectors. The eigenvectors for each of the two
matrices need not be orthogonal.
\noindent
The eigenvalue spectrum $\{\gamma \}$ of the matrix $\mbox{\boldmath $A$}\mbox
{\boldmath $B$}$ (or
$\mbox{\boldmath $B$}\mbox{\boldmath $A$}$) determines the stability of the
exact solutions. If $\{\gamma \}$ contains negative eigenvalues then the solution is 
unstable. The eigenvalue spectrum always contains two eigenvalues which are 
zero. These eigenvalues correspond to the translational invariance in space
and time (represented by $\delta_1$ and $\delta$).
The given solutions are unstable for most of the parameter space
$(\zeta, \xi, \nu)$. From Ref. \cite{krss1} we know that $\xi=0$ generally leads to stable
solutions.
%\begin{figure}[h]
%\includegraphics[width=0.5\textwidth]{Fig_param.ps}
% \caption{\label{fig:Parameter_region}Illustration of stability equivalence in the parameter space. It is argued (see text for details)
%that the light (dark) grey regions have equivalent stability properties which can be fully determined on the 
%thick solid (dashed) line segments.}
% \end{figure}
In determining the stability of the solutions it is useful to note 
that the rescaling transformation $(\zeta, \xi, \nu) \rightarrow \alpha
(\zeta, \xi, \nu)$ ($\omega\rightarrow\alpha\omega$) changes the eigenvalues 
by $\alpha^2$ and therefore it does not affect the stability (i.e. the sign of 
eigenvalues) of the solutions. Therefore the 
three-dimensional parameter space $(\zeta, \xi, \nu)$ can be significantly
reduced (into a two-dimensional parameter space) as far as stability considerations are
concerned. The nonlinearity parameter $\nu \neq 0$, separates
the three-dimensional parameter space into two disconnected equivalent ones for 
$\nu > 0$ and $\nu<0$, respectively.
The rescaling transformation with $\alpha=-1$ interchanges the two equivalent
half spaces. So we need only to consider positive $\nu$. From here on we treat the
symmetric and the asymmetric cases separately.

{\noindent \it Stability of the symmetric case:}

Since $\nu > 0$, Eqs. (\ref{dn symm cond}), (\ref{cn symm cond}),
and (\ref{pulse symm cond})
require $\zeta + \xi > 0$ and therefore
 $\zeta > -\xi$. Further, we can 
always choose $\zeta > \xi$  due to the inter-changeability of $\zeta$ and
$\xi$.  We therefore have $-\zeta < \xi < \zeta$ or applying the scaling 
condition $-1< \xi/\zeta <1$.  This means that the stability of the entire
parameter space can be mapped out onto the much smaller parameter space
$(1,\xi, \nu)$, where $-1< \xi < 1$ (and $\nu > 0$). 
\begin{figure}[h]
\includegraphics[width=0.5\textwidth]{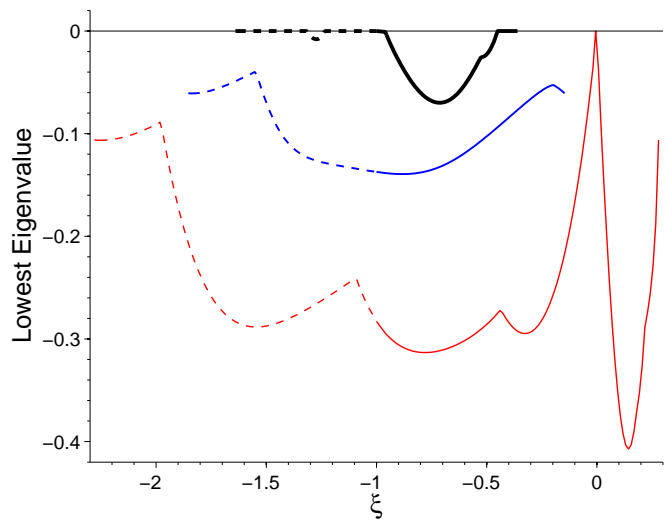}
 \caption{\label{fig:sol}Stability analysis for the $dn$ solution,  
Eq. (\ref{dn symm}). Lowest eigenvalues for $|\nu|=1.5$ (thick line, black online), 
$|\nu|=2$ (medium thick line, blue online), and $|\nu|=3$ (thin line, red online). Dashed (solid)
lines indicate negative (positive) values of $\nu$. Stability occurs when the
lowest eigenvalue is zero. 
The entire existence interval is shown for positive $\nu$. Negative $\nu$  
results can be obtained by rescaling the results for positive $\nu$. The remaining
parameters are: $\zeta=1$ and $N_p=8$.}
\end{figure}
\begin{figure}[h]
\includegraphics[width=0.49\textwidth]{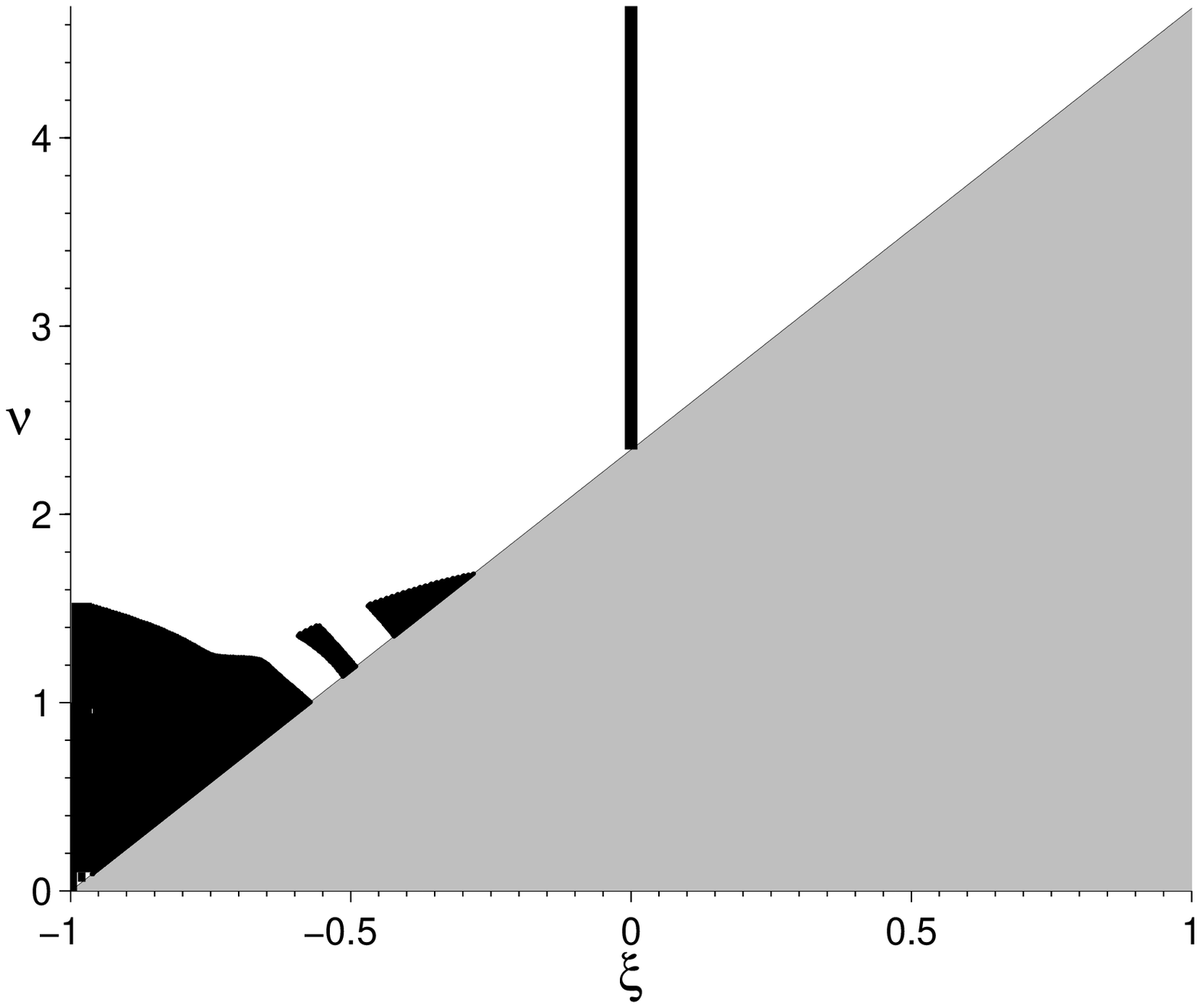}
\includegraphics[width=0.49\textwidth]{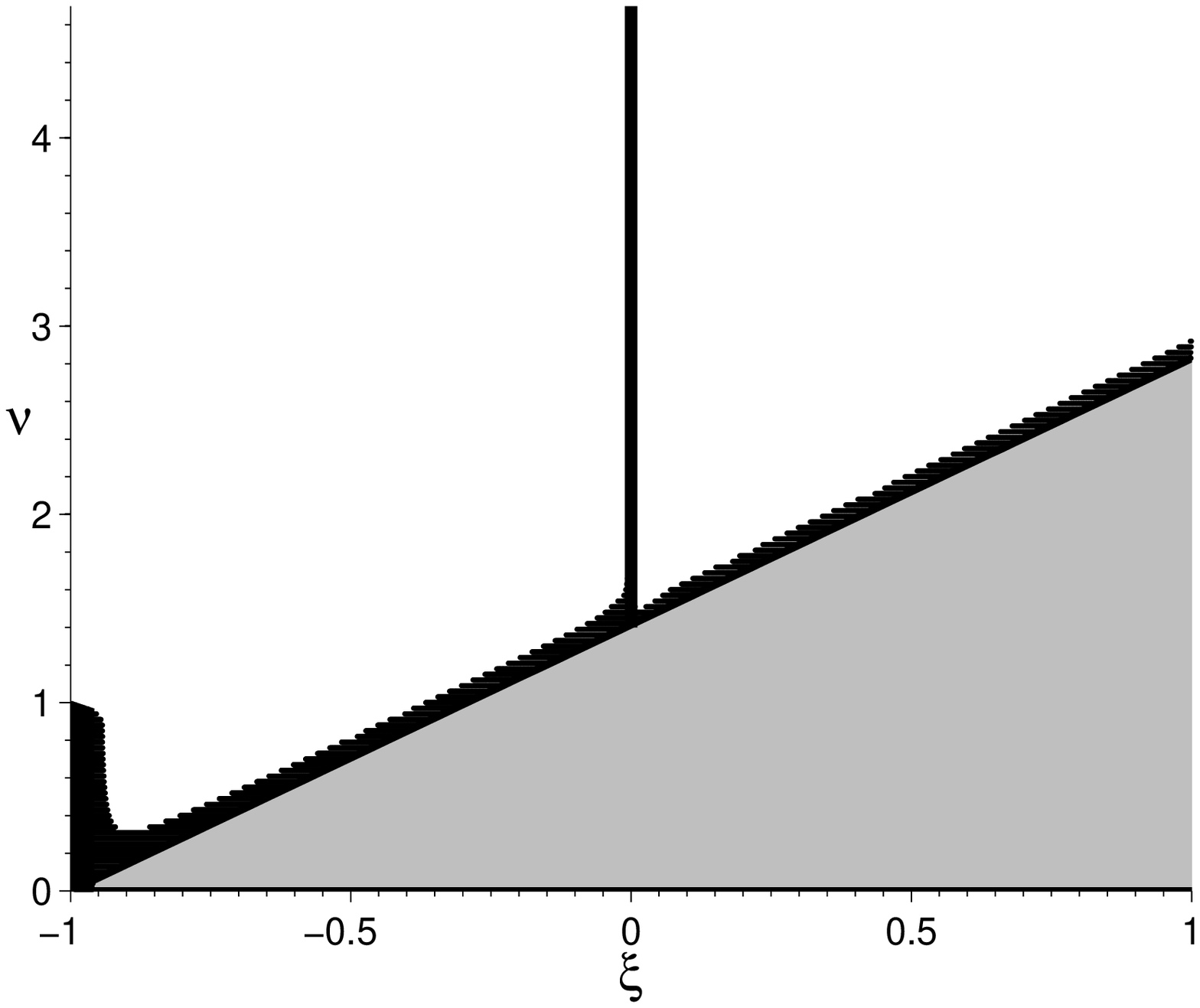}
 \caption{\label{fig:symmetric}General stability diagram for symmetric $dn$ (left) and $cn$ (right) 
solutions of Eq. (\ref{dn symm}) and Eq. (\ref{cn symm}), respectively. In the grey region no solution 
of the given kind exists.  In the white regions solutions are unstable, while the black area indicates
stable solutions. The vertical line $\xi=0$, which represents the one-dimensional stability result, is 
part of the black area.  The rough or jagged appearance of parts of the stability boundary is due to 
limited numerical  resolution rather than an intrinsic feature of the  problem. Parameters are: 
$\zeta=1$, and $N_p=8$.}
 \end{figure}
In Fig. \ref{fig:sol} we illustrate the stability analysis for the $dn$ solution, 
Eq. (\ref{dn symm}), by showing the lowest
eigenvalue as a function of $\xi$ for several 
values of $\nu$. Stability occurs whenever the lowest eigenvalue is zero.
The entire existence regime is illustrated 
for positive $\nu$ and a few windows of stability can be seen. It is important
to note that the results for negative values of $\nu$ ($\xi<-1$) are
superfluous as they can be obtained by rescaling the results for positive $\nu$.
To demonstrate this we have for $\zeta=1$: $(1,-|\xi|,-|\nu|)\rightarrow
(-1/|\xi|,1,|\nu|/|\xi|)\rightarrow (1,-1/|\xi|,|\nu|/|\xi|)$ where the 
last step follows by the 
inter-changeability of the two-coupling parameters. This shows that the 
$\nu <0$ 
regime can be mapped onto the $\nu>0$ regime. Generally, stability is 
only observed for $\xi <0$. This means that stability for the symmetric $\dn$
solutions can only be achieved when $\xi$ and $\zeta$ have opposite signs. 
Recalling the equivalence between staggered solutions and sign changes of
$\xi$ or $\zeta$, another way to view this result is that 
the symmetric $\dn$ solutions must be staggered in one dimension 
when $\zeta$ and $\xi$ are both positive in order to be stable. This is clearly a 
property arising from the discreteness that cannot be achieved in a continuum system.

Assembling results like those shown in Fig. \ref{fig:sol} for a range 
of $\nu$ and $\xi$ values
we arrive at Fig. \ref{fig:symmetric} where stability
diagrams for both the $dn$ and the $cn$ solutions to Eq. (\ref{dn symm}) and
Eq. (\ref{cn symm}), respectively, are shown.
The grey region indicates that the solutions do not
exist, whereas the black (white) regions indicate the existence of stable
(unstable) solutions. These two stability diagrams have a very 
similar structure. However, the $cn$ solutions are always 
stable in the proximity of the existence boundary marked by the grey area.
This property is related to the fact that the amplitude (which is $\propto k$,
see Eq.(\ref{cn symm})) of this solution vanishes at the existence 
boundary where $k=0$. Also, we note that
the common pulse solution corresponds to the corner close to $(\xi,\nu)=(-1,0)$.

We have looked at other values of $N_p$ and find
that for larger $N_p$, the stability diagram has similar features. The pulse-like solution 
only exists in the limit $N_p \rightarrow \infty$, and here it has the same stability 
properties as the $dn$ and $cn$ solutions for $k \rightarrow 1$. Therefore, our analysis 
indicates that the symmetric pulse-like solutions are stable for small $\nu$ and $\xi \sim -1$. 

%\begin{figure}
%\includegraphics[width=0.5\textwidth]{stability_cn.ps}
% \caption{\label{fig:stacn}General stability diagram for symmetric $cn$ solution
%of Eq. (10). In the grey region no solution of the given kind exists.
%In the white region solutions are unstable. While the black area indicate
%stable solutions. This include the vertical line $\xi=0$, which represents the
%one-dimensional stability result. Parameters are: $\zeta=1$, and $N_p=8$.}
% \end{figure}

{\noindent \it Stability of the asymmetric case:}
\begin{figure}[h]
\includegraphics[width=0.49\textwidth]{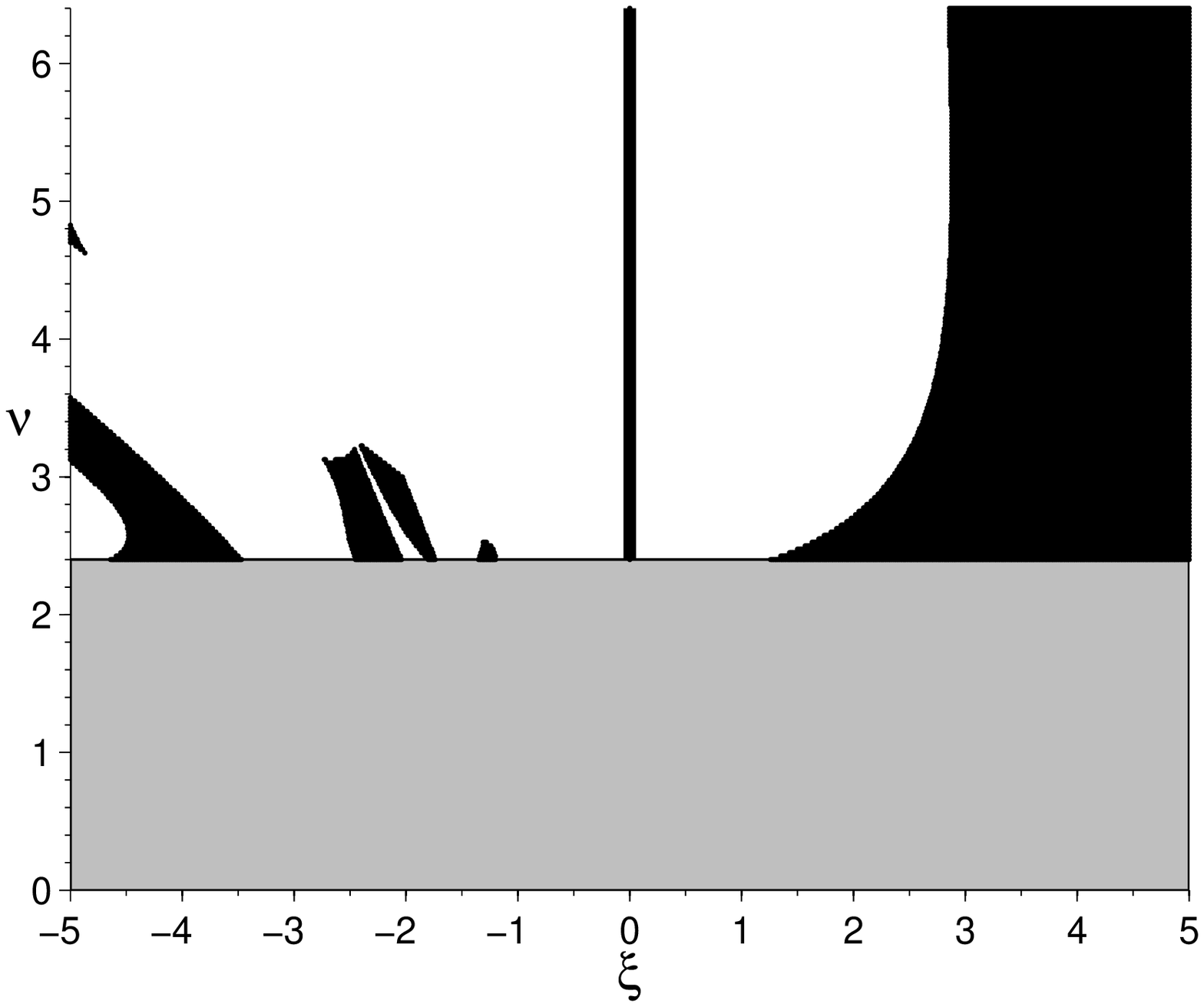}
\includegraphics[width=0.49\textwidth]{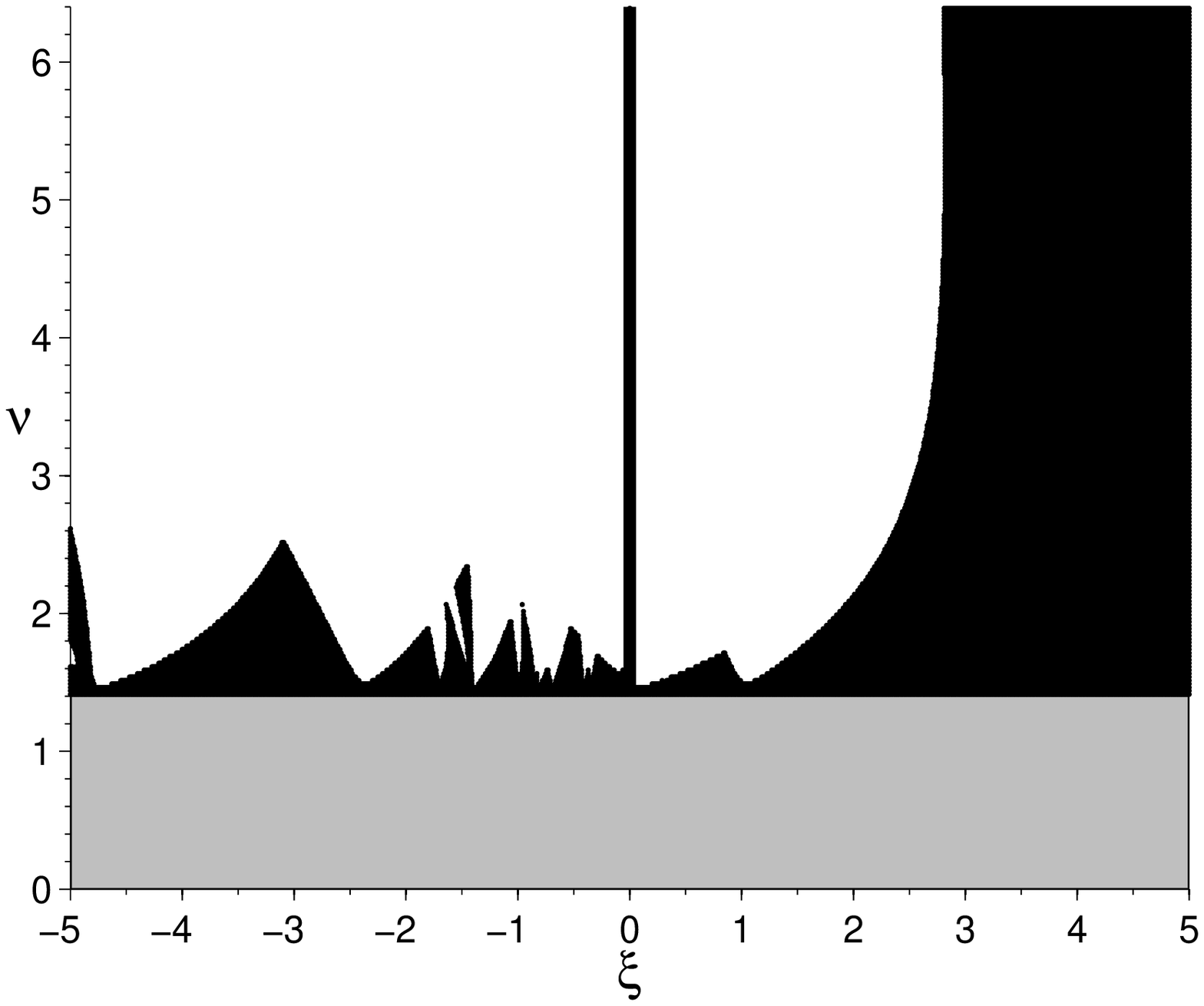}
 \caption{\label{fig:asymmetric}General stability diagram for the 
asymmetric $dn$ (left) and $cn$ (right) 
solutions of Eq. (\ref{dn asymm}) and Eq. (\ref{cn asymm}), respectively. 
In the grey region no solution of the given 
kind exists.  In the white region solutions are unstable, while the black 
area indicates stable solutions. The vertical line $\xi=0$, which represents
the one-dimensional stability result, is part of the black area. 
Parameters are: $\zeta=1$, and $N_p=8$. }
 \end{figure}

%\begin{figure}
%\includegraphics[width=0.5\textwidth]{Stability_Asym_dn.ps}
% \caption{\label{fig:staAdn}General stability diagram for asymmetric $dn$
%solution of Eq. (15). In the grey region no solution of the given kind exists.
%In the white regions solutions are unstable. While the black area indicate
%stable solutions. This include the vertical line $\xi=0$, which represents the
%one-dimensional stability result. Parameters are: $\zeta=1$, and $N_p=8$.}
% \end{figure}
%\begin{figure}
%\includegraphics[width=0.5\textwidth]{Stability_Asym_cn.ps}
% \caption{\label{fig:staAcn}General stability diagram for asymmetric $cn$
%solution of Eq. (17). In the grey region no solution of the given kind exists.
%In the white regions solutions are unstable. While the black area indicate
%stable solutions. This include the vertical line $\xi=0$, which represents the
%one-dimensional stability result. Parameters are: $\zeta=1$, and $N_p=8$.}
%\end{figure}
We proceed almost as in the symmetric case. We still have $\nu>0$, therefore
from Eqs. (\ref{dn asymm cond}), (\ref{cn asymm cond}), and
(\ref{pulse asymm cond}) we have $\zeta>0$.
So here the three-dimensional parameter space can be reduced to
$(1,\xi,\nu)$, where $-\infty<\xi<\infty$ (and $\nu>0$).
Illustrations of the stability diagrams are given for the asymmetric $dn$
and $cn$ solutions, Eq. (\ref{dn asymm}) and Eq. (\ref{cn asymm}),
respectively in Fig. \ref{fig:asymmetric} for $\zeta=1$ and $N_p=8$.
Again the two stability diagrams have a very similar structure except that 
the parameter space for stability of the asymmetric solution is much larger.
As in the symmetric case, even in the asymmetric case, the $cn$ solutions 
are always stable in the 
proximity of the existence boundary marked by the grey area. 
Here we note that
the common pulse solution corresponds to large $\nu$ 
and $\xi$.
% \stackrel{_>}{_\sim} 2.8$.

\section{\bf Peierls-Nabarro barrier for the pulse solution} 

We would now like to show the absence of Peierls-Nabarro barrier
for the pulse solution.  However, we must remember that since both power $P$ and
Hamiltonian $H$ are constants of motion, one must compute the energy difference
between the solutions when $\delta_1=0$ and $\delta_1=1/2$ in such a way that 
the power $P$ is {\it same} in both the cases. 

For the pulse solution obtained above, the power $P$ is given by
\be\label{h4}
P=\sum_{n,m=-\infty}^{\infty} |\phi_{n,m}|^2 =\sinh^2(\beta)
\sum_{n,m=-\infty}^{\infty} \sech^2 [\beta(n+m+\delta_1)]\,.
\ee

This double sum can be evaluated using the single sum result
\be\label{h5}
\sum_{n=-\infty}^{\infty} \sech^2 [\beta(n+\delta_2)] =\frac{2}{\beta}
+\frac{2K(k)E(k)}{\beta^2}+\left(\frac{2K(k)}{\beta}\right)^2\dn^2[2\delta_2 K(k),k]\,,
\ee
the above $P$ is given by
\be\label{h6}
P=\sum_{m=-\infty}^{\infty} \sinh^2(\beta) \left[\frac{2}{\beta}
+\frac{2K(k)E(k)}{\beta^2}+\left(\frac{2K(k)}{\beta}\right)^2
\dn^2[2(m+\delta_1) K(k),k] \right]\,.
\ee
Note, only the last term on the rhs is $m$ dependent. In these 
equations $E(k)$ is the complete elliptic integral of the second kind.

Let us now discuss the computation of the Hamiltonian $H$. Clearly, for
the pulse solution obtained above, $H$ as given by Eq. (\ref{Hamiltonian}) 
takes the form
\bea\label{h7} 
&&H=\sum_{n,m=-\infty}^{\infty} \bigg [-\nu P
+\nu
\ln[1+\sinh^2(\beta)\sech^2(\beta[n+m+\delta_1])] \nonumber \\
&&-2\sinh^2(\beta) [\sech(\beta[n+1+m+\delta_1])
\sech(\beta[n+m+\delta_1])(\zeta+\xi)] \bigg ] \,.
\eea
Again we use the single sum results to evaluate the double sum, i.e.
\be\label{h8}
\sum_{n=-\infty}^{\infty} \big [\sech[\beta(n+1+\delta_2)] 
\sech[\beta(n+\delta_2)] \big ]=\frac{2}{\sinh(\beta)}\,,
\ee
\be\label{h9}
\sum_{n=-\infty}^{\infty} \ln[1+\sinh^2(\beta)\sech^2(\beta[n+\delta_2])]
=2\beta\,,
\ee 
the above $H$ is given by
\be\label{h10}
H=\sum_{m=-\infty}^{\infty} \bigg [-\nu P
+2 \nu \beta- 4(\zeta+\xi)\sinh(\beta) \bigg ]\,.
\ee
We thus note that for a given power $P$ (which contains a sum over $m$), H is 
indeed independent of  $\delta_1$, i.e. the Peierls-Nabarro barrier is indeed zero for the pulse 
solution.  The same holds true for the asymmetric solution.

\section{\bf Short period solutions} 

Recently we obtained short period solutions to the one-dimensional
saturable DNLSE  \cite{krss2}. 
These short period ($N$) solutions, in the one-dimensional case, can be written
in the following compact form (coming from equally distributed points on
a circle so that projection on the $x$-axis should only be 0 or $\pm a$):
\be
u_N(n)=\frac{a}{\cos(\varphi_N)}\cos\left(\frac{2\pi n}{N}+\varphi_N\right),
\ee
where
$\varphi_1$
=$\varphi_2$ 
=$\varphi_{4s}=0$, 
(4s is the stable period 4) and 
$\varphi_4=\frac{\pi}{4}$,
$\varphi_3=\varphi_6=\frac{\pi}{6}$.
\bea
u_N(n+1)+u_N(n-1)=
\frac{a}{\cos(\varphi_N)}\bigg[\cos\left(\frac{2\pi(n+1)}{N}+\varphi_N\right)+
\cos\left(\frac{2\pi(n-1)}{N}+\varphi_N\right)\bigg]\\
=\frac{2a}{\cos(\varphi_N)}\cos\left(\frac{2\pi n}{N}+\varphi_N\right)
\cos\frac{2\pi}{N}
=2\psi_N(n)\cos\frac{2\pi}{N}.
\eea
For $\omega$ we get
\be
\omega= -2\zeta\cos\frac{2\pi}{N}-\frac{\nu a^2}{1+a^2}.
\ee

Assuming that in the two-dimensional case, the solution is a product of 
the two one-dimensional solutions (with period $N$ and period $M$), i.e.
\be
u_{N,M}(n,m)=\frac{a}{\cos(\varphi_N)\cos(\varphi_M)}
\cos\left(\frac{2\pi n}{N}+\varphi_N\right)
\cos\left(\frac{2\pi m}{M}+\varphi_M\right),
\ee
%\[
%\bigg(=a\frac{\cos\big(2\pi(\frac{n}{N}+\frac{m}{M})
%+\varphi_N +\varphi_M)
%+\cos(2\pi(\frac{n}{N}-\frac{m}{M})
%+\varphi_N -\varphi_M\big)}
%{\cos(\varphi_N+\varphi_M)+
%\cos(\varphi_N-\varphi_M)}\bigg)
%\nonumber
%\]
we get:
\be
\omega= -2\zeta\cos\frac{2\pi}{N}
-2\xi\cos\frac{2\pi}{M}
-\frac{\nu a^2}{1+a^2}.
\ee

\section{\bf Conclusions}

We have given analytical expressions for the solutions to the two-dimensional
discrete nonlinear Schr{\"o}dinger equation with saturable nonlinearity which 
arises in photorefractive crystals \cite{efrem, fleischer, martin, melvin, magnus}. 
Due to their infinite extension along one of their dimensions, these solutions are 
not very physically meaningful but it is very rare that solutions to discrete nonlinear two-dimensional problems 
can be described in closed form using standard mathematical functions as we 
have done here. This feature of the solutions is physically significant 
because it provides an opportunity for in-depth scrutiny and understanding
that is not usually available in a nonlinear physical system.  
These solutions are closely related to the previously derived \cite{krss1} solutions to the
corresponding one-dimensional equation. However, in contrast to what one may
expect based on intuition derived from similar nonlinear partial differential
equations, we have shown that these solutions are linearly stable in certain
regions of the parameter space.  Specifically, we have observed that the 
asymmetric versions of these solutions lead to a very intricate stability diagram. 
We have shown that the symmetric $dn$ solution is stable in certain regions of the 
parameter space provided it is staggered in one dimension. However, the symmetric 
$cn$ solution as well as the asymmetric $cn$ and $dn$ solutions are stable in certain 
regions of parameter space both when they are non-staggered or if they are staggered 
in one dimension. The finding that nonlinear waveforms in two-dimensional photorefractive materials best achieve
stability in the presence of phase asymmetry between the two spatial directions is crucial because 
the photonic lattices that represent the physical realization of Eq.(\ref{field EQ}) tend to naturally possess this 
property \cite{OL}.
Finally, we found that the Peierls-Nabarro barrier for  the pulse solution 
is zero.  An understanding of the mobility of these exact discrete two-dimensional
solutions remains an important issue \cite{magnus}. 

\acknowledgments
This work was carried out under the auspices of the National Nuclear Security
Administration of the U.S. Department of Energy at Los Alamos National
Laboratory under Contract No. DE-AC52-06NA25396.

\end{document}